\documentclass{aa}
\usepackage{graphicx}
\usepackage{epsfig}
\def\ltsima{$\; \buildrel < \over \sim \;$}
\def\gtsima{$\; \buildrel > \over \sim \;$}
\def\simlt{\lower.5ex\hbox{\ltsima}}
\def\simgt{\lower.5ex\hbox{\gtsima}}

\begin{document}

   \title{A changing inner radius in the accretion disc of Q0056--363? }

   \author{G. Matt\inst{1}, D. Porquet\inst{2}, S. Bianchi\inst{1,3}, S. Falocco\inst{1}, 
R. Maiolino\inst{4}, J. N. Reeves\inst{5} \and L. Zappacosta\inst{5}
          }

   \offprints{G. Matt}

   \institute{Dipartimento di Fisica, Universit\`a degli Studi Roma Tre,
via della Vasca Navale 84, I-00146 Roma, Italy
         \and
Max-Planck-Institut f\"{u}r extraterrestrische Physik, Postfach 1312, 85741 Garching, Germany
\and 
XMM-Newton Science Operation Center, ESAC/ESA, Apartado 50727, E--28080 Madrid, Spain 
\and
INAF--Osservatorio Astrofisico di Arcetri, Largo Fermi 5, 50125 
Firenze, Italy
\and
Laboratory for High Energy Astrophysics,  NASA/GSFC, 
Greenbelt Road, Greenbelt, MD 20771, U.S.A.
\and 
Department of Physics and Astronomy, University of California, Irvine, 4129
Frederick Reines Hall, Irvine, CA 92697-4575, U.S.A.
}

   \date{Received ; Accepted }

   \abstract{ Q0056-363 is the most powerful X-ray quasar known to exhibit a broad,
likely relativistic iron line (Porquet \& Reeves 2003). It has been observed twice
by XMM--$Newton$, three and half years apart 
(July 2000 and December 2003). In the second observation, the UV and soft
X--ray fluxes were fainter, the hard X--ray power law flatter, and the iron line equivalent
width (EW) smaller than in the 2000 observation. 
These variations can all be explained, at least qualitatively, if the
disc is truncated in the second observation. We report also on the possible detection of a
transient, redshifted iron absorption line during the 2003 observation.
   }

\authorrunning{G. Matt et al.}
\titlerunning{A changing inner radius in the accretion disc of Q0056--363? }

   \maketitle
%
%________________________________________________________________

\section{Introduction}

The geometrical and physical properties of the inner regions of the accretion discs around
the supermassive Black Holes in Active Galactic Nuclei are still largely unknown, despite
many theoretical and observational efforts. While there is wide consensus 
that Comptonization of thermal disc emission by hot electrons is the main mechanism
of production of X--rays, the details of this process are by no means clear. The geometry
of the inner accretion disc regions, and their very existence, are also
matter of debate. While in a few sources relativistic iron line profiles are clearly
observed (e.g. Fabian et al. 2002, Turner et al. 2002), 
implying that the accretion disc is probably extending down to the innermost stable orbit
(i.e. 1.23$r_g$ if $a$=1,  or 6$r_g$ if $a$=0, where 
$r_g=GM/c^2$ and $a$ is the Black Hole angular
momentum per unit mass), in many sources only ``narrow'' (i.e. unresolved)
iron components are present (e.g. Bianchi et al. 2004, and references therein). 
These narrow lines probably originate in distant matter, like
the BLR or the torus, even if in many cases an origin in the outer regions of the accretion
disc cannot at present be ruled out. 

The reason why in many sources there is no evidence for the 
 presence of the innermost regions of the accretion disc
is unclear. One possible explanation is ionization of the matter. If the
matter is mildly (or very highly) ionized, iron line emission is efficiently suppressed  
(Matt et al. 1996). Of course, ionization is expected to be higher in the innermost regions,
so mainly suppressing the broader components of the iron line. 
Another possibility is disc truncation. In Galactic Black Hole systems, disc truncation 
is believed
to play an essential role in explaining the spectral differences between Soft and Hard states, 
the main physical parameter likely being the accretion rate 
(e.g. Done \& Gierlinski 2004; Fender
et al. 2004, and references therein). In the Hard state, the disc is truncated
while, in the Soft state, it extends down to the innermost stable orbit. An
Intermediate state occurs during the transitions between the two states. 
The disc emptying and refilling is associated with matter ejection, as signalled by radio
emission. The radio properties in fact
also changes dramatically from state to state. In the Hard state, there is evidence for a steady
jet, emitting strongly at radio wavelenghts via synchrotron radiation. Matter ejection 
is quenched in the High state, while in the Intermediate state the ejection may be present,
but often in form of discrete blobs.  
The disc emptying and refilling may occur on very different time scales, from seconds 
(as in GRS~1915+105, Belloni et al. 1997, where  probably only a limited part of the cycle,
i.e. between Intermediate and High states, is covered: Fender et al. 2004) to months. 
A similar situation may well occur in AGN, and indeed Marscher et al. (2002) 
have found that the superluminal Seyfert 1 galaxy 3C120, when observed simultaneously
in radio and X--rays, behaves similarly to  GRS~1915+105, even if, as obvious, on longer
time scales. Disc truncation has also been invoked by Zdziarski et al. (1999) 
and Lubinski \& Zdziarski (2001) to explain the observed correlation
between the X-ray spectral slope and both the amount of Compton
reflection and the EW (and width) of the iron line. In fact, if the disc is truncated, 
the strength of the reprocessed features (iron line and Compton reflection continuum)
are reduced because of the smaller solid angle subtended by the matter to the primary
X-ray source; the line width is also expected to decrease, because the Doppler and
Gravitational broadenings are most relevant at small radii. The amount of thermal
disc emission is also reduced, and the hard X--ray spectrum is flattened
due to the less effective cooling. 
It is worth noting that these relations between $\Gamma$ (the power
law spectral index) and the reprocessed features are not expected in 
the ionization scenario, where the reduction of the line EW due to
resonant trappings or full matter ionization is not accompanied by a reduction of the
thermal disc emission and therefore of the cooling in the Comptonizing corona.

In this paper we present possible evidence for a changing inner radius in the
broad line radio--quiet quasar Q0056-363 ($z$=0.162). The source
 was observed by XMM--$Newton$ twice. In the first observation, performed on July 2000 and 
about 20 ks long, an intense (EW of 
about 250 eV) and broad (velocity width of about 25000 km s$^{-1}$) iron line was present,
strongly suggesting emission from the innermost regions of the accretion disc (Porquet \&
Reeves 2003). Q0056-363 is presently the most luminous quasar with an observed relativistic line.
The source was then reobserved on December 21-23, 2003 for about 90 ks. Here we report
on the latter observation, and the comparison between the two.

\begin{figure*}
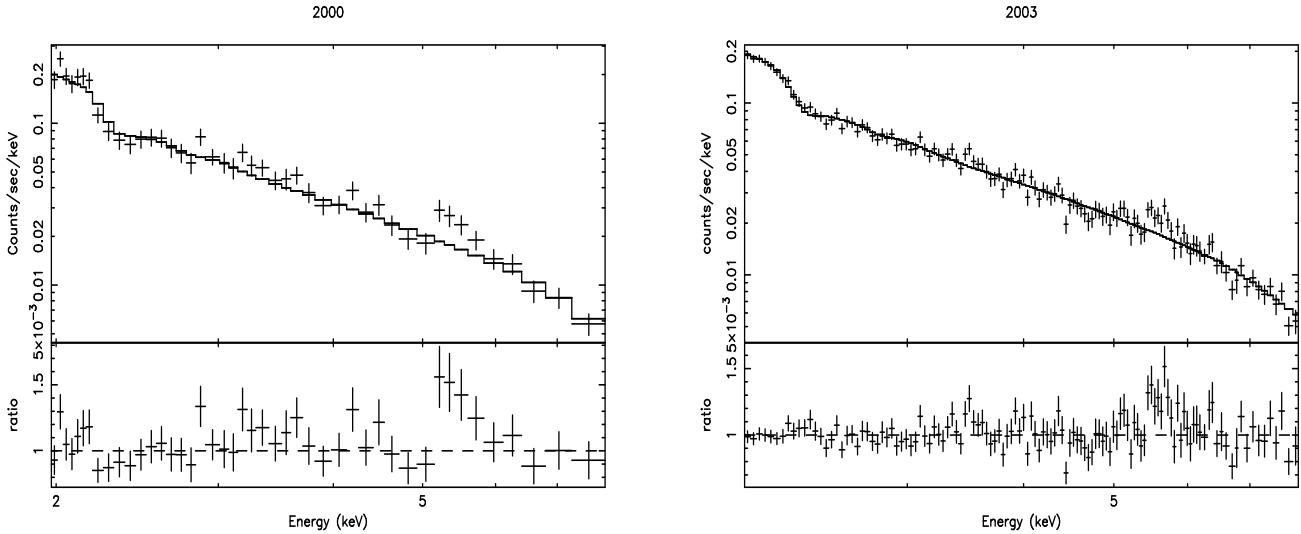

\hbox{
\epsfig{figure=line2000.ps,height=8.0cm,width=7.0cm,angle=-90}
\hspace{1.0cm}
\epsfig{figure=line2003.ps,height=8.0cm,width=7.0cm,angle=-90}
}
\caption{
{{\it Left panel}: data, best fit model and data/model ratio in the iron line region
for the 2000 observation, when fitted with a simple power law.  
{\it Right panel}: the same, but for the 2003 observation.}}
\label{lines}
\end{figure*}

\begin{table*}[t]
\begin{center}
\begin{tabular}{|c|c|c|c|c|c|c|c|}
\hline
& ~ & ~ & ~ & ~ & ~ & ~  & ~ \cr
~ & FLUX (UWV2) & FLUX (0.5-2 keV) & FLUX (2-10 keV) & 
$\Gamma_{\rm soft}$ & $\Gamma_{\rm hard}$ & $\sigma_{\rm K\alpha}$ &  EW$_{\rm K\alpha}$ \cr
~ & (erg cm$^{-2}$ s$^{-1}$ A$^{-1}$) 
& (erg cm$^{-2}$ s$^{-1}$)  & (erg cm$^{-2}$ s$^{-1}$)
 & ~ & ~ & (eV) & (eV) \cr 
& ~ & ~ & ~ & ~ & ~ & ~ & ~ \cr
\hline
& ~ & ~ & ~ & ~ & ~ & ~ & ~ \cr
2000 & 2.11$\times$10$^{-14}$  & 3.40$\times$10$^{-12}$ & 2.96$\times$10$^{-12}$ & 
3.35$^{+0.46}_{-0.14}$ & 1.87$^{+0.17}_{-0.06}$ & 270$^{+180}_{-130}$ & 
266$^{+134}_{-99}$ \cr
& ~ & ~ & ~ & ~ & ~ & ~ & ~\cr
2003 & 1.35$\times$10$^{-14}$ & 3.02$\times$10$^{-12}$ & 2.90$\times$10$^{-12}$ & 
2.87$^{+0.02}_{-0.02}$ & 1.61$^{+0.03}_{-0.02}$ & 250$^{+100}_{-100}$ &
118$^{+38}_{-40}$  \cr
& ~ & ~ & ~ & ~ & ~ & ~ & ~ \cr
\hline
\end{tabular}
\caption{The best fit parameters for the 2000 and 2003 observations. The model consists 
of two power laws and a gaussian line centred at 6.4 keV (rest frame).}
\label{fits}
\end{center}
\end{table*}

\section{Data reduction and analysis}

Both XMM-\textit{Newton} observations of Q0056-363 were reduced with
\textsc{SAS} 6.0.0, following standard procedures and screening periods of
high background flaring according to the method presented in Piconcelli et
al. (2004), based on the cumulative distribution function of background
lightcurve count rates. This method, which maximes the signal-to-noise
ratio, resulted in net exposure times of $\simeq14$ ks and $\simeq94$ ks,
for the 2001 and the 2003 observation, respectively. EPIC PN (Str\"uder et
al. 2001) and MOS (Turner et al. 2001) spectra were extracted, adopting
circular extraction regions with radii of 40 and 45 arcsec, respectively.
After inspecting the results from the \textsc{SAS} tool \textsc{Epatplot}
and from some fits on single and double pattern spectra separately, we
concluded that the effect of pileup is negligible in both observations,
allowing us to use pattern 0-4 and 0-12 for the PN and the MOS. 
Due to operational
needs, the first MOS observation was splitted in two $\simeq5$ ks segments,
with different filters. In both observations, the addition of the MOS 
in the spectral fits do not significantly increase the precision with which
spectral parameters are determined. For these reasons, and for the sake of simplicity,
we will use and discuss in the following only results from the PN, unless explicitely
stated.

For spectral fitting, the spectra were rebinned in order to have at least 25 counts per bin
and to oversample the energy resolution by at least a factor of 3.
Spectral fits are performed with the software package {\sc xspec}.
All reported errors will correspond to 90\% confidence level for one
interesting parameter ($\Delta\chi^2$=2.71).

\section{Results}

\subsection{Comparison between the 2000 and 2003 observations}

In both observations, a simple absorbed power law (with Galactic absorption fixed at
1.94$\times10^{20}$ 
cm$^{-2}$)\footnote{\sc http://heasarc.gsfc.nasa.gov/cgi-bin/Tools/w3nh/w3nh.pl} 
fails to fit the data, while the inclusion of a soft excess
and an iron line are sufficient to reasonably model the spectra. 
For the sake of simplicity,
and because we are interested here mainly to the comparison between the 2000 and 2003 spectra,
rather than to any detailed modeling, 
we parametrized the soft excess with a simple power law, and the iron line with a gaussian with 
centroid energy fixed at 6.4 keV (rest frame) and $\sigma$ as a free parameter.
This model provides good enough fits over the 0.5-12 keV energy band
($\chi^2$/d.o.f. of  127.3/127 and 359.4/223 for the 2000 and 2003 
observations, respectively). The best fit parameters are reported in Table~\ref{fits}. 

While the full band fluxes in the two observations are rather similar, spectral changes
are apparent. In the 2000 observation, both the soft and hard power laws are significantly 
steeper than in the 2003 observation, while the iron line is more intense (even if at the
90\% confidence level the EWs are  only just inconsistent each other). The width of the line
is instead poorly constrained. 
 In Fig.~\ref{lines}, the 2--10 keV spectra and residuals,
when fitting with a simple power law, are shown. An eye inspection would tell
that the line in the 2000 is broader, but the quality of the data, especially for the 2000
observation, is not good enough to make this difference statistically significant.

Similar results are obtained leaving the line centroid energy free to vary. The centroid
energy, $\sigma$ and EW are: 6.31$^{+0.19}_{-0.10}$ keV, 270$^{+130}_{-100}$ eV, 
240$^{+112}_{-80}$ eV (2000), and  6.47$^{+0.09}_{-0.09}$ keV, 220$^{+100}_{-80}$ eV, 
109$^{+39}_{-32}$ eV (2003). The statistical quality of the fits did not improve. 

If the line is fitted with a relativistic profile in Schwarzschild metric, 
with fixed centroid energy (6.4 keV) outer radius (1000$r_g$), 
inclination angle (30$^{\circ}$), and emissivity law (power law with index --2),
the best fit inner radius is 37$^{+47}_{-29}$ $r_g$ for the
2003 observation, and basically unconstrained ($r<100r_g$) for the 2000 observation. In both 
cases, the quality of the fit is similar to those with a gaussian line.  Again, the results
did not change significantly leaving the centroid energy free to vary.

\begin{figure}
\epsfig{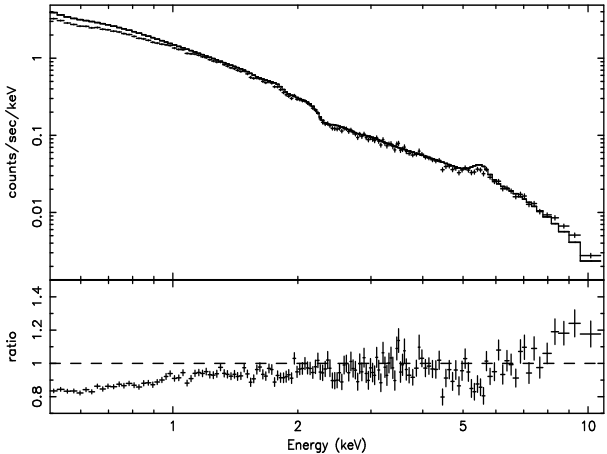}
\caption{The 
2003 spectrum when fitted with the 2000 best fit model. Note the harder spectrum,
both at low and high energies, and a deficit at the iron line energies (5.5 keV in
the observed frame), due to the fact that in the 2000 observation the line is more intense.
}
\label{comparison}
\end{figure}

To illustrate the spectral changes, in Fig.~\ref{comparison} we show the 2003 spectrum 
and residuals when fitted with the 2000 best fit model (no renormalization is made).
An excess of photons at high energies, and an even more 
dramatic deficit at low energies are indeed
apparent, along with a deficit at the iron line energies.  In the soft band (0.5-2 keV)
the flux in 2000 is higher than in 2003 by about 13\%, while in the hard band (2-10 keV)
the difference is very small, only 3\%. 
The UV flux, measured with the OM, also changed between the two observations,
being about 1.6 times larger in 2000 (see Table~\ref{fits}). All these numbers indicate
flux variations increasing in amplitude with decreasing photon frequency.

\subsection{Time--resolved analysis of the 2003 observation}

We also searched for short term spectral changes, by dividing the 2003 observation
in 5 intervals. No statistically significant changes, 
either in the continuum or line properties, were found.  

\begin{figure}
\epsfig{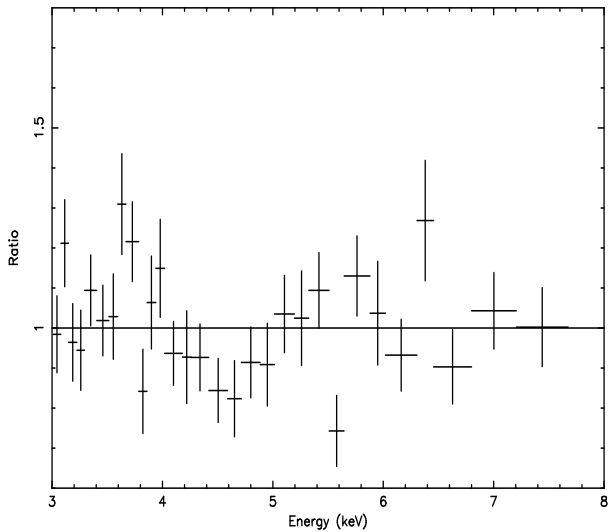}
\caption{The spectrum of the third interval in the 
2003 observation. An absorption feature at $\sim$4.5 keV (observer frame) is apparent.}
\label{abs}
\end{figure}

We also searched for transient emission and absorption features, as those recently
observed in many sources (see e.g.  Porquet et al 2004; Dov\v{c}iak et al. 2004, and references
therein). We set, rather arbitrarily, a confidence level of 99\% (according to the F-test)
as the detection threshold for a feature. An 
absorption feature in the third interval has been found to meet this criterion
(see Fig.~\ref{abs}). If fitted with a gaussian
with $\sigma$ fixed to 0.1, the centroid energy is 5.34$^{+0.12}_{-0.13}$ keV
(rest frame) and
the EW is --75$^{+34}_{-41}$ eV. The significance of this feature is 99.2\%.
Leaving $\sigma$ free to vary, we do not obtain a significantly better fit. 
In the MOS, this feature is neither required nor ruled out. As a word
of caution, we note that a possible emission line
at about 3.7 keV (observed frame) could also be present, but having  
a significance of 98.7\% is just below our threshold.

%-----------------------------Figure Start--------------------------------

\section{Discussion}

The spectral changes described in the previous section
are consistent with a `truncating disc'  
scenario in which the disc inner radius was closer to the BH  
in 2000 while the disc was truncated in 2003. 
The smaller soft excess, lower UV flux, flatter hard X-ray spectrum,
fainter iron line in the latter observation
can all be explained 
in this way, at least qualitatively. In fact, in a truncated disc the thermal
emission (likely responsible for the UV and soft X--ray emission) is obviously smaller,
the iron line is less intense (due to the smaller solid angle subtended by the disc to the
primary X--ray emitting
region) and less broad (because the relativistic effects are less important),
and the hard X--ray spectrum is flatter due to the less effective cooling.  
A strong prediction of this scenario is  
a narrower iron line in the 2003 observation. Unfortunately, the quality
of the spectra is not good enough to draw any conclusion  in this respect, and for
the time being disc truncation must be considered only a plausible and appealing
hypothesis. Giving its potential interest, however, we discuss this possibility
more extensively.

Disc truncation has been proposed by Zdziarski et al. (1999) 
to explain the observed correlation between
the strength of the Compton Reflection component
and the spectral slope in AGN and Galactic Black Hole systems, and by
Lubinski \& Zdziarski (2001) to explain the  correlation between
the iron line width and EW  and the slope.

In Galactic Black Hole systems, disc emptying and refilling is believed to play
an important role, and this phenomenon seems to be strongly related to matter ejection 
(e.g. Fender  et al. 2004, and references therein). In fact  
ejection of matter is usually associated with Hard X-ray states both at 
relatively low and very high accretion rates (e.g. Fender  et al. 2004).
In the Low/Hard state, where $\dot{m}$ is $\simlt$1\% the Eddington rate,
the ejection is in form of a steady jet (Gallo et al. 2003), and the innermost regions of the
accretion disc is absent. In the High/Soft state (high $\dot{m}$), the disc extends down to the 
innermost stable orbit; radio emission is suppressed, suggesting the quenching 
 of matter ejection. During the transition from Low to High states, there is a
phase (called Very High State) in which the disc is refilling, the hard (i.e. 
RXTE--PCA band) X--ray luminosity is quite stable at large values 
but with still a hard spectrum (continuously softening 
until the High state is reached; see Fig.~7 of Fender et al. 2004), and there are episodes of
high velocity ejections. It is also worth noting that some
sources (most notably GRS~1915+105, Fender \& Belloni 2004 and references therein)
switches between High and Very High states only (with corresponding variations of the inner
disc radius) without going through the entire cycle. 
If this scenario is, at least qualitatively, applicable
also to AGN (as suggested by the scaling of the radio/X-ray coupling across a range of
$\simgt$10$^7$ solar masses, Merloni et al. 2003, Maccarone et al. 2003), 
and given the fact that $\dot{m}$
seems to be large in both observations (Porquet \& Reeves 2003 
estimated $L/L_{\rm edd} \sim$0.6-0.8,
and in the 2003 observation the hard X-ray flux is similar), 
one can speculate that the source
was in a High State in 2000, and in a Very High State in 2003. 

In this scenario, we can also fit the absorption feature at about 5.3 keV (rest frame)
possibly detected in the spectrum of the third interval in which we
divided the 2003 observation. The most natural interpretation
is in term of a redshifted iron line. If entirely due to Doppler effect, this implies
$v$=0.2$c$ if the line is the He--like, or $v$=0.23$c$ if the line is the H--like one. 
If instead the redshift is entirely gravitational, this implies (in Schwarzschild metric)
a location of
the absorbing matter at $r\sim$5.5 or 4.9 $r_g$ (assuming that the primary X--rays are
produced very close to be blob, so that there is no shift between the two regions).
The equivalent width is pretty large, implying
a large column density (at least 10$^{23}$ cm$^{-2}$) and 
either a significant iron overabundance or a strong contribution from turbulence (Bianchi 
et al. 2005). 

Iron redshifted absorption features have possibly been already observed in a 
few objects  (e.g. Nandra et al. 1999; Longinotti et al. 2003). Interestingly, Dadina et al.
(2005) have found evidence for transient redshifted iron lines in the 
BeppoSAX spectrum of Mrk~509, both red-- and blue--shifted. It is worth noting that transient and
shifted absorption lines are naturally expected in models involving blobby ejection and 
downfalling of matter,
as in the recently proposed aborted--jet model of Ghisellini et al. (2004), where at least
part of the heating of the X--ray emitting corona is due to collision by blobs ejected with
a velocity smaller than the escape velocity. Absorption by one of these blobs, either
downfalling (the $\sim$20 ks observing time corresponds,
for a velocity of about 0.2$c$ and a Black Hole of about 5$\times$10$^8$ M$_{\odot}$, to 
a crossing distance of about a Schwarzschild radius)
or very close to the BH, and then gravitationally redshifted,
could explain our findings.
If, as in Galactic Black Hole systems, ejection
of matter preferentially occurs when the accretion disc is truncated, it is not surprising
to find such a feature in the 2003 observation.

\section{Conclusions}

Comparing the 2000 and 2003 observations of the quasar, Q0056-363, 
we have found differences which can be interpreted as
a change of the disc inner radius, which was possibly smaller (i.e. closer to the Black Hole
event horizon) in the former observation. In the latter observation we have also found 
evidence, even if marginal,
for a transient absorption line, interpreted as redshifted ionized iron.  Intriguingly,
these findings can be qualitatively fitted into the scenario proposed for stellar mass Black Hole
systems (Fender et al. 2004), thus possibly
providing a link between properties of accreting
Black Holes accross a range of masses of several orders of magnitude.

\section*{Acknowledgements}

We thank Elena Gallo for enlightening discussions, and the anonymous referee for useful
comments. 
GM acknowledges financial support from MIUR under grant {\sc prin-03-02-23}.
This paper is 
based on observations obtained with XMM-Newton, an ESA science
mission with instruments and contributions directly
funded by ESA Member States and the USA (NASA).

\end{document}